\documentclass[10pt,a4paper]{article}

    \usepackage[T1]{fontenc}
    \usepackage{inputenc}
    \usepackage[english]{babel}
    \usepackage[nohead,includefoot,margin=2cm]{geometry}
    \usepackage[compact,raggedright,small]{titlesec}
    \usepackage[affil-it]{authblk}
    \usepackage[runin]{abstract}
    \usepackage{multicol}
    \usepackage{blindtext}
\usepackage{subfig}
\usepackage{graphicx}
\usepackage{cite}

    \newcommand*{\pacsname}{PACS numbers}

    \addto{\Affilfont}{\small}

    \title{\large\bfseries An Accurate Circuit Model for Photoconductive Antennas}

    \author[1]{Fatemeh aghili}
    \author[1,*]{Ali Mahjoori}
    \affil[1]{Department of Electrical and Computer Engineering, University of Sistan and Baluchestan,  Iran, 98167-45845}
\affil[*]{Corresponding author: al.mahjoori@gmail.com}
    
\begin{document}
       \date{}
       \maketitle

       \begin{abstract}
THz photoconductive antennas are among the most popular devices for THz wave generation and detection. A novel and accurate circuit is proposed to model the performance of such antennas, especially those which are frequency dependant. The suggested equivalent circuit consists of an external source voltage, a source input admittance, a loss resistance related to the electrode loss resistance, a frequency-dependant capacitance associated with capacitive  behaviour of the gap between the electrodes, and a controlled voltage-dependant source representing the screening of photo-carriers. Based on this model, the radiation from a simple frequency dependent antenna, i.e. a dipole antenna, is investigated. Excellent agreement is observed between the obtained results by the proposed equivalent circuit model and those by measurement.
       \end{abstract}

{T}{erahertz} (THz) band is defined as the region of the electromagnetic spectrum ranging from 0.1 THz to 10 THz which is between the millimeter and far-infrared frequencies \cite{1}. Very recently, the interest in this band of frequency has increased enormously because of its various potential applications such as sensing, medical imaging, and extremely broadband communication \cite{2,3,4,5,6,7,8,9,10,11,12,13}.\\
\indent To realize THz generation and detection several approaches are demonstrated, among which THz photoconductive antennas have been of significant attention in established THz experimental setups \cite{14,15}. A typical THz photoconductive antenna is mainly composed of two electrodes which are placed on top of an ultrafast photoconductive substrate (usually LTG-GaAs). An optical laser first illuminates the gap between the electrodes which produces free carriers. These carriers are then accelerated by a bias voltage applied between the electrodes, which results in producing a photocurrent following the optical pump envelope. The generated photocurrent is fed to the electrodes to radiate THz wave into free space \cite{1}.\\
\indent An appealing approach to analyze the performance of THz photoconductive antennas is to model its radiation characteristics via an equivalent circuit model. To this end, four distinct equivalent circuits have already been proposed. The first model included two constant resistance representing the antenna input impedance and the associated loss of the photoconductive material, and a constant capacitance related to the gap capacitance of the electrodes \cite{16}. Despite the simplicity of this model, it was not accurate enough to address all aspects of the radiation of the antenna such as screening of the photo-carrier. The second equivalent circuit model was then proposed to improve the accuracy of the previous one \cite{17}. The circuit was composed of a constant resistance related to the antenna input impedance, a time-varying source resistance representing the screening of the photo-carriers. Although this model was much more accurate than the first one, the complexity in the derivation of its elements was significantly high. The third equivalent circuit model was presented to fill out the gap between previously mentioned models. This equivalent circuit model not only considered the screening effect of the photo-carrier, but also was highly straight. The equivalent circuit consisted of several lumped elements whose values were derived based on the complex physical mechanisms \cite{18}. There were still however some important factors influencing the radiation of the antenna which were neglected in the model, like the temperature and the operation frequency of the antenna. The latest model was developed so as to address the temperature dependance of the radiation of the antenna \cite{19}. However, neither of these models consider the effect of operational frequency of the antenna. In fact, while there exist many important and appealing frequency dependant antennas such as a simple dipole one, the aforementioned models are only appropriate for frequency-independent antennas like a bow-tie one. \\
\indent In this paper, our aim is to develop an accurate and comprehensive  circuit model to model photoconductive antennas including those having dependency to the operation frequency. To this end, we start with considering the frequency domain counterpart of the equations previously used to analyze photoconductive antennas in time domain. We then develop an accurate and efficient model for them. Finally, we confirm the validity of the model by obtaining the radiation characteristics of a simple dipole antenna and comparing them with those obtained by measurement.\\
 \begin{figure}[!h]
  \centering
  \includegraphics[width=0.45\textwidth]{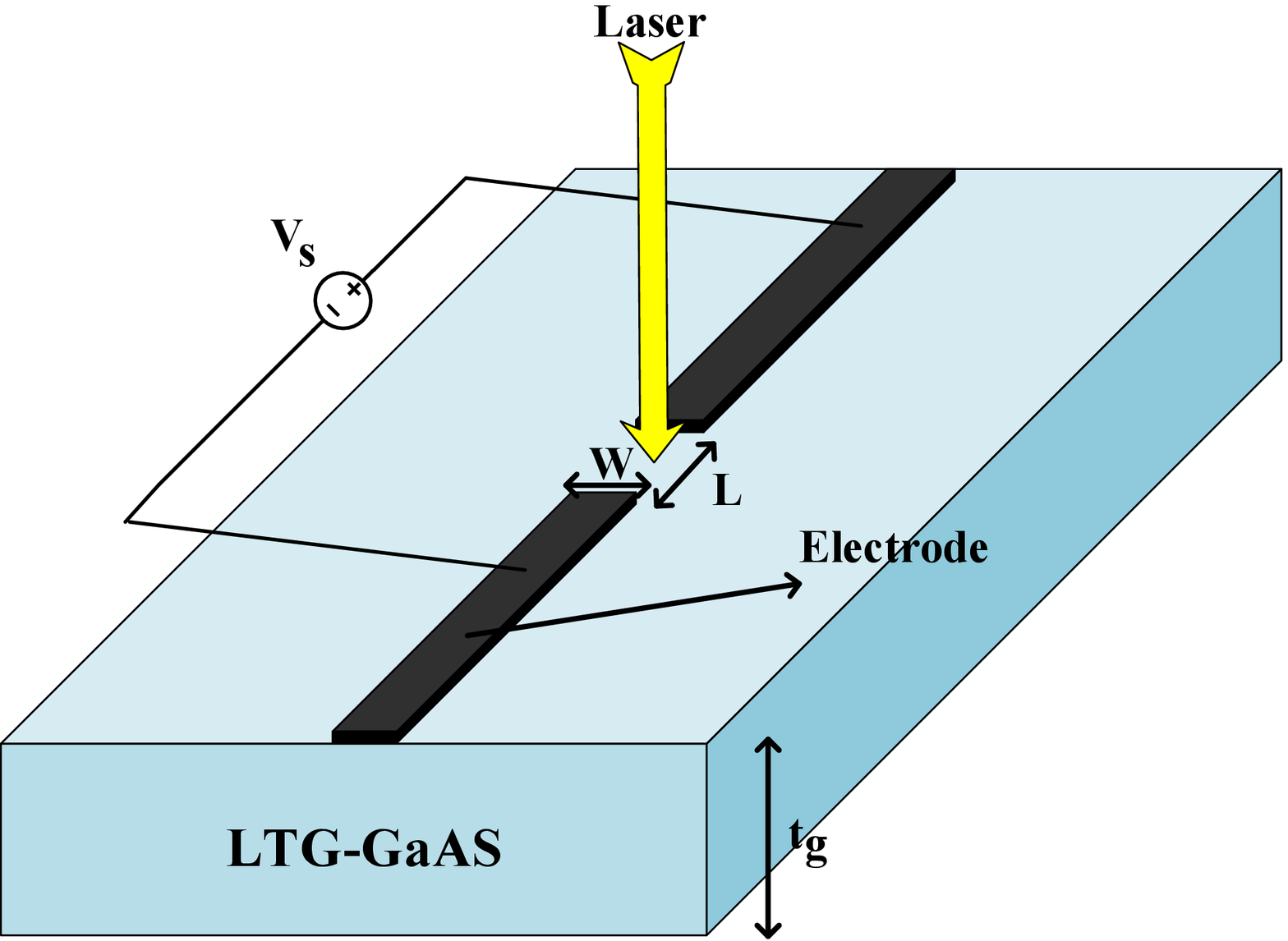}\ \center(a)\\
  \includegraphics[width=0.45\textwidth]{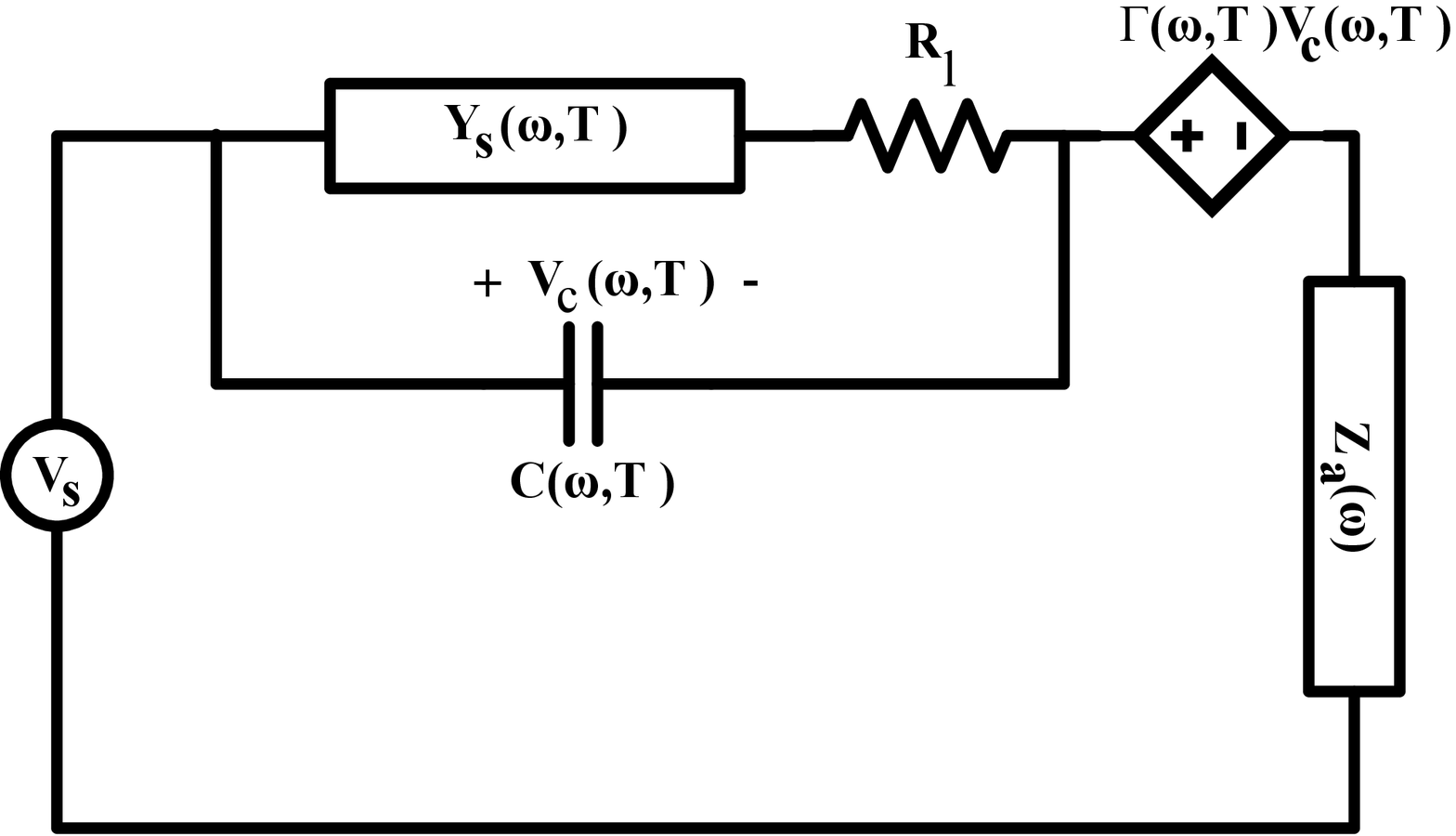} \center (b)
  \caption{(a) Schematic diagram of a typical THz dipole photoconductive antenna. (b) Equivalent circuit of the THz photoconductive antenna.}\label{fig1}
\end{figure}
The schematic diagram of a typical THz photoconductive antenna is depicted  in Fig. 1(a), which includes all major elements explained before. Since the gap size of the antenna is much smaller than the wavelength of
THz band, it is possible to model such device by an equivalent lumped-element circuit \cite{20} whose elements are frequency dependent. Fig. 1(b) shows our proposed equivalent circuit model for the photoconductive antenna. The model includes the following components:\\
\indent 1) A sours voltage $V_{s}$ associated with the external applied voltage.\\
\indent 2) A frequency dependant source admittance $Y_{s}(\omega)$  representing the behaviors of the laser pulses and the photo-carriers. To drive the parameter $Y_{s}(\omega)$, we first calculate the carrier density $N(\omega,T)$ generated by the illumination using Drude model given by \cite{21}
\begin{equation}\label{1}
N(\omega,T)=\frac{1}{(j\omega+\frac{1}{\tau_{c}})}\times\frac{\alpha(T)}{h\nu_{opt}}I_{l}(r,\omega)
\end{equation}
where $T$ is temperature, $\tau_{c}$ is the carrier lifetime, $h$ is the Planck's constant, $\nu_{opt}$ is the laser frequency, $I_{l}(r,\omega)$ is the optical pulse intensity and $\alpha(T)$ is the optical absorption coefficient defined as \cite{22}
\begin{equation}\label{2}
 \alpha(T)\approx K_{abs}\sqrt{\frac{h\nu_{opt}-Eg(T)}{q}}
\end{equation}
in which $K_{abs}$ is a definite constant which is about $9.7\times10^{15}$ for GaAs \cite{20}, $\nu_{opt}$ is the laser frequency and $Eg$ is band gap energy of GaAs which is expressed as follows \cite{23}
\begin{equation}\label{eq3}
  E_{g}(T)=E_{g}(0)-\frac{\alpha_{E}T^{2}}{T+\beta_{E}}
\end{equation}
where  $E_{g}(0)$ is the energy gap at $0~K$  which is about $1.519~eV$ for GaAS, $\alpha$ and $\beta$ are material-specific constant which are about $5.41\times10^{-4}~\frac{eV}{K}$ and $204~K$ for GaAs, respectively \cite{23}.\\
Here, we assume the electric field distribution of the laser pulses to be Gaussian. The optical pulse density $I_{l}(r,\omega)$ can then be expressed as
\begin{equation}\label{2}
I(r,\omega)=\int_{-\infty}^{+\infty}I_{l}(1-R)exp(-\frac{-2r^{2}}{w_{0}^2})exp(-\frac{2t^{2}}{\tau_{l}^2})exp(-j\omega t)dt
\end{equation}
in which $I_{l}$ is the laser peak intensity, $R$ and $w_{0}$ are power reflection coefficient and beam waist radius, respectively, and $\tau_{l}$ is the laser pulse duration. Making use of Eqs. 1 and 4 $N(\omega,T)$ can be written as

{\setlength\arraycolsep{2pt}
\begin{eqnarray}
N(\omega,T)=\frac{1}{(j\omega+\frac{1}{\tau_{c}})}\times\frac{\alpha(T)}{h\nu_{opt}}\times ~~~~~~~~~~~ \nonumber \\ I_{l}(1-R)exp(-\frac{2r^{2}}{w_{0}^{2}})exp(-\frac{(\omega-2\pi \nu_{opt})^{2}}{2\tau_{l}^{2}})
\end{eqnarray}}
Assuming that the mobility of electrons is much higher than the mobility of holes, the source admittance of the antenna will be obtained as
{\setlength\arraycolsep{2pt}
\begin{eqnarray}
Y_{s}(\omega,T)=\int dY_{s}(\omega,T)=\int_{0}^{t_{g}}\sigma_{s}(\omega,T)exp(-\alpha(T) z)\frac{W}{L}dz\nonumber\\
=\frac{W}{L\alpha(T)}\sigma_{s}(\omega,T)(1-exp(-\alpha(T) t_{g}))~~~~~~~~~~~~
\end{eqnarray}}
 where $L$, $W$ and $t_{g}$ are antenna gap length, width and thickness, respectively, and $\sigma_{s}(\omega,T)$ is of the form
\begin{equation}\label{4}
\sigma_{s}(\omega,T)=e\mu_{e}(T)N(\omega,T)
\end{equation}
where $e$ is electron charge and $\mu_{e}(T)$ is the mobility of electrons obtained based on Einstein equation as
\begin{equation}\label{1}
\frac{D}{\mu_{e}(T)}=\frac{kT}{e}
\end{equation}
in which $D$  and $k$ are diffusion and  Boltzman constant. By substituting  Eqs. 2 and 7 into Eq. 6, the value of the source admittance can easily be calculated. \\
\indent 3) A resistance $R_{loss}$ taking into account the loss resistance of the electrodes and photoconductive material. \\
\indent 4) A frequency-dependant capacitance $C(\omega,T)$ corresponding to the screening  effect of the photo-carriers near the electrodes. We will obtain an accurate expression for $C(\omega,T)$ in the following of the paper. \\
5) A frequency-dependant voltage-controlled source controlled by the voltage across the capacitance, i.e. $\Gamma(\omega,T)V_{c}(\omega,T)$.  We will also obtain the accurate expression of $\Gamma(\omega,T)$ in the following.\\
\indent 6)The antenna impedance, $Z_{a}(\omega)$ which is assumed to have dependency to frequency. \\
\indent We next focus our attention to find the values of proposed parameters $C(\omega,T)$ and  $\Gamma(\omega,T)$. Applying circuit analysis in the circuit shown in Fig. 1(b), one obtains
\begin{equation}\label{3}
I(\omega,T)=V_{c}(\omega,T)G_{s}(\omega,T)+2j\omega C(\omega,T)V_{c}(\omega,T)
\end{equation}
Using Eq. 9, the gap voltage $V_{c}(\omega,T)$ can be easily obtained as
\begin{table}
\caption{Laser, Photoconductive Material And Antenna Parameters .}

\label{TableI}
\begin{center}
\begin{tabular}{c|c|c}
\hline
               &         &         \\
    Parameter  & notation &  Value   \\
               &          &         \\

\hline
Reflection Coefficient  & $R$ & 0.318  \\
Laser frequency \cite{16} & $\nu_{opt}$ &$ 384~ THz$     \\
Laser repetition rate \cite{16}& $f_{rep}$ &$ 82~MHz $ \\
Laser pulse duration \cite{16} & $\tau_{l}$ &$ 80~fs $ \\
Carrier lifetime \cite{9} & $\tau_{c}$ &$ 1~ps $ \\
Carrier recombination time \cite{9} & $\tau_{r}$ &$ 100~ps $ \\
Screening factor \cite{9}& $\zeta$ &$900$ \\
Antenna gap length \cite{16} & $L$& $ 5~\mu m $ \\
Antenna gap width \cite{16} & $W$ &$ 10~\mu m $ \\
Depth of excitation area & $t_{g}$ &$1~\mu m$ \\
Energy gap of GaAs at $0~K$ \cite{14} & $E_{g}(0)$ &$1.519~eV$ \\
Diffusion constant of GaAS & $D$ & $200 \frac{cm^{2}}{s}$\\
\hline
\end{tabular}
\end{center}
\end{table}
{\setlength\arraycolsep{2pt}
\begin{eqnarray}
V_{c}(\omega,T)=\frac{1}{2j\omega}\times \{ {\frac{V_{s}(\omega)}{Z_{a}(\omega)C(\omega,T)}}- {\frac{V_{c}(\omega,T)}{Z_{a}(\omega)C(\omega,T)}} \nonumber \\
-\frac{\Gamma(\omega,T)V_{c}(\omega,T)}{Z_{a}(\omega)C(\omega,T)}-\frac{G_{s}(\omega,T)V_{c}(\omega,T)}{C(\omega,T)} \}
\end{eqnarray}}
 We then match  Eq. 7 to the physical carrier transport behavior of the photoconductive antenna when illuminated by laser pulses. The electric field distribution in the antenna gap is formulated in  \cite{24} as

{\setlength\arraycolsep{2pt}
\begin{eqnarray}
j\omega E_{c}(\omega,T)=\frac{1}{K(\omega,T)}\{\frac{E_{s}(\omega)}{\tau_{r}}-\frac{1}{\tau_{r}}E_{c}(\omega,T)\nonumber \\
-\frac{e\mu_{e}(T)N(\omega,T)E_{c}(\omega,T)}{\varepsilon\zeta} \nonumber \\
-\frac{e\mu_{e}(T)Z_{a}(\omega)SN(\omega,T)E_{c}(\omega,T)}{L\tau_{r}}\nonumber \\
-\frac{e\mu_{e}(T)Z_{a}(\omega)SE_{c}(\omega,T)}{L}j\omega N(\omega,T) \}
\end{eqnarray}}
in which $\tau_{r}$ is the recombination lifetime, $\zeta$ is the geometrical factor of the substrate \cite{9}, $\varepsilon$ is substrate permittivity and $K(\omega,T)$ is defined as
\begin{equation}\label{6}
K(\omega,T)=1+\frac{e\mu_{e}(T)Z_{a}(\omega)SN(\omega,T)}{L}
\end{equation}
Comparing Eq. 10 and Eq. 11, the values of $C(\omega,T)$ and $\Gamma(\omega,T)$ will then be of the form
\begin{equation}\label{4}
C(\omega,T)=\frac{\tau_{r}}{Z_{a}(\omega)}(1+\frac{e\mu_{e}(T)Z_{a}(\omega)SN(\omega,T)}{L})
\end{equation}
\begin{equation}\label{5}
\Gamma(\omega,T)=\frac{e\mu_{e}(T)N(\omega,T)\tau_{r}}{\xi \varepsilon}
\end{equation}
\begin{figure}
  \centering
  \includegraphics[width=0.5\textwidth]{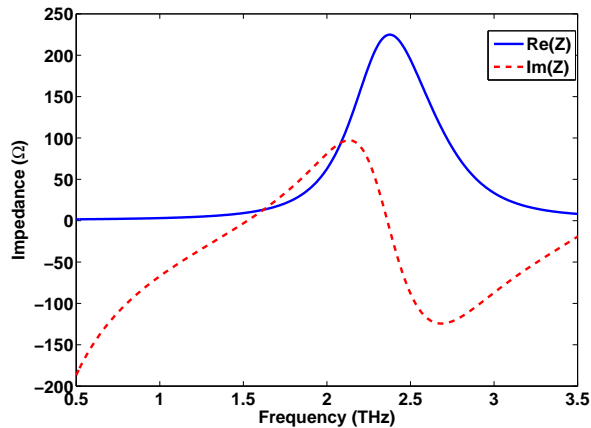}
  \caption{Simulated real (solid) and imaginary (dashed) part of input impedance of the investigated photoconductive antenna.}\label{fig5}
\end{figure}
Once the parameters $\Gamma(\omega,T)$ and $C(\omega,T)$ are calculated making use of Eqs. 14 and 15, the equivalent circuit of the photoconductive antenna will be accomplished.\\
 \indent It is worth mentioning that some  simplified assumptions have been applied during modeling process. Fist of all, it is assumed that the electric field distribution in the gap is uniform. Moreover, according to \cite{24}, it is presumed that the carrier relaxation time, and consequently the carrier mobility of photo-carriers are small. Therefore, the proposed circuit model will maintain adequate if the substrate of the antenna has a small carrier mobility.\\
 In order to analyze  the behavior of photoconductive antennas based on the introduced equivalent circuit, the associated parameters of a typical THz photoconductive antenna is summarized in Table I. The antenna input impedance $Z_{a}(\omega)$ of a simple diploe with length $L=30~ \mu m$ and width $W=10~ \mu m$ has been calculated using full-wave simulations, and is depicted in Fig. 2. Once the input impedance $Z_{a}(\omega)$ is found, the induced photocurrent at the antenna gap can easily be determined based on the abovementioned equivalent circuit.\\
\indent  According to basics of the THz photoconductive antennas \cite{21}, the induced photocurrent at the antenna gap can be expressed as $I_{pc}(\omega,T)=eN(\omega,T)v(\omega,T)$, in which $v(\omega,T)$ is the average carrier velocity in the antenna gap. The average carrier velocity is defined as $v(\omega,T)=\mu(T)E_{c}(\omega,T)$, where $E_{c}(\omega,T)$ is the electric field in the the antenna gap, which in turn is defined as  $E_{c}(\omega,T)=\frac{V_{c}(\omega,T)}{L}$. Therefore, the photocurrent at the antenna gap can be written of the form
\begin{equation}\label{j}
I_{pc}(\omega,T)=eN(\omega,T)\mu(T)V_{c}(\omega,T)\frac{S}{L}
\end{equation}
in which $V_{c}(\omega,T)$ is the antenna gap voltage obtained based on Eq. 10. Once the induced photocurrent in frequency domain is calculated, the temporal behaviour of the photocurrent can simply be obtained by taking inverse Fourier transform. Fig. 3 illustrates the temporal behaviour of the obtained THz photocurrent at the antenna gap. The calculated photocurrent based on the proposed equivalent circuit is in great agreement with the photocurrent measured in \cite{25} for a similar dipole antenna, as the half-width of the produced pulse is $0.2~ps$ in both cases.
\begin{figure}
  \centering
  \includegraphics[width=0.5\textwidth]{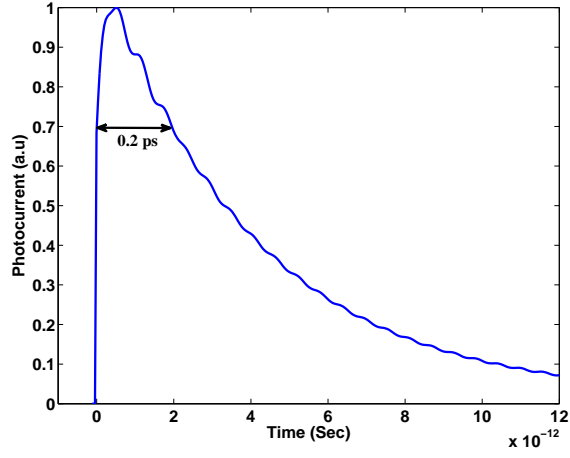}
  \caption{Temporal behaviour of the THz photocurrent at the antenna gap.}\label{fig5}
\end{figure}
\indent  The radiated THz field of the photoconductive antenna is then investigated. From the basics of the THz photoconductive antennas, the radiated field of a THz photoconductive antenna is proportional to the  derivative of the THz photocurrent at the antenna gap \cite{1}. Therefore, we have
\begin{equation}\label{2}
E_{THz}(\omega,T)=j\omega I_{pc}(\omega,T)
\end{equation}
As expressed, the temporal behaviour of the radiated field can be obtained by applying an inverse Fourier transform. Fig. 4 illustrates the temporal behaviour of the radiated field of the proposed dipole antenna. Excellent agreement is observed between the obtained result and the result in \cite{24}, confirming the accuracy of the proposed model.
\begin{figure}
  \centering
  \includegraphics[width=0.5\textwidth]{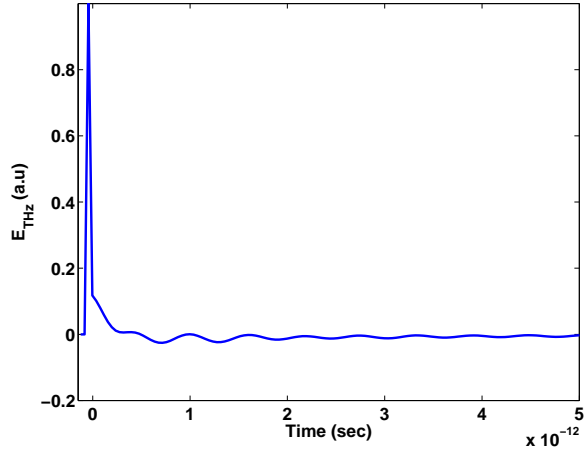}
  \caption{Temporal behaviour of the THz radiated field.}\label{fig5}
\end{figure}
\\
In this paper, we proposed a novel equivalent circuit model for frequency dependant THz photoconductive antennas. The proposed equivalent circuit was composed of an external source voltage, a source input admittance, a loss resistance regarding the electrode loss resistance, a frequency-dependant capacitance representing capacitive behaviour of the gap between the electrodes, and finally a controlled voltage-dependant source representing the screening of photo-carriers. The elements of the circuit were determined, one by one. The radiation from a simple frequency dependent dipole antenna was then investigated based on the proposed equivalent circuit model, and compared with the results carried out by measurements. Excellent agreement was observed between these two results confirming the accuracy of the model.

\bibliography{sample}

    \end{document}